# Deep learning for fast MR imaging: a review for learning reconstruction from incomplete k-space data

Shanshan Wang, Taohui Xiao, Qiegen Liu, and Hairong Zheng

*Abstract*—Magnetic resonance imaging is a powerful imaging modality that can provide versatile information but it has a bottleneck problem "slow imaging speed". Reducing the scanned measurements can accelerate MR imaging with the aid of powerful reconstruction methods, which have evolved from linear analytic models to nonlinear iterative ones. The emerging trend in this area is replacing human-defined signal models with that learned from data. Specifically, from 2016, deep learning has been incorporated into the fast MR imaging task, which draws valuable prior knowledge from big datasets to facilitate accurate MR image reconstruction from limited measurements. This survey aims to review deep learning based MR image reconstruction works from 2016- June 2020 and will discuss merits, limitations and challenges associated with such methods. Last but not least, this paper will provide a starting point for researchers interested in contributing to this field by pointing out good tutorial resources, state-of-the-art open-source codes and meaningful data sources.

*Index Terms*—**Deep learning, MRI, undersampled image reconstruction**

## I. Introduction

MAGNETIC resonance imaging (MRI) is a non-invasive imaging technique that can provide rich anatomical and functional information. Nevertheless, MRI has a bottleneck problem "slow imaging speed" a.k.a "long imaging time", which consists of two parts, i.e. acquisition and reconstruction. To accelerate MR scan, three mainstream methods have been developed, namely, physics based fast imaging sequences, hardware based parallel imaging with multiple coils and signal processing based MR image reconstruction from incomplete k-space data. These techniques along with their combinations have either reduced the acquisition or reconstruction time. Our specific focus here is the signal processing based methods, which explore prior knowledge to regularize the image reconstructions from reduced measurements.

During the past several decades, MR image reconstructions have evolved from linear analytic methods to nonlinear iterative ones. Typical examples in analytic reconstruction consist of partial Fourier, sensitivity encoding (SENSE) [1], SMASH [2], and GRAPPA [3], which explore the prior knowledge of K-space correlations and the sampling properties of the imaging system. These methods have been extensively used in commercialized scanners based on their high efficiencies. On the other hand, the nonlinear methods focus on iteratively solving a model that considers both the physics of the imaging system and the prior knowledge of the object being imaged. Popular prior includes sparsity [4], low-rank [5]–[8], statistics distribution regularization [9], generalized series (GS) model [10], manifold fitting [11]–[13] and so on so forth [14], [15]. These prior knowledges have been mostly developed via model or hand designs, such as dictionary learning [16] and total variation [17], [18]. These methods have achieved great successes in their episodes, with commercialized available MR scanners. However, there are still limitations with these two mainstream methods. For the first phase methods, normally no object prior is considered and it suffered from relatively long scanning time. For the second phase methods, mainly compressed sensing (CS) techniques, the prior model capacity is limited. The weighting parameter of the CS models are hard to tune. Furthermore, the image prior model capacity is still limited. The emerging trend in this area is replacing human-defined signal models with that learned from data.

Deep learning, a technique which is driving today's artificial intelligence (AI) explosion, uses neural networks with many layers of processing units to learn complex patterns in large amount of data [19]. It has attracted unprecedented public attention, showing great potential for diverse medical imaging tasks [20]–[25]. Its popularity is driven with the advances in computing power and improved training techniques. In 2016, deep learning has been incorporated into the fast MR imaging task for the first time, which draws valuable prior knowledge from big datasets to facilitate accurate MR image reconstruction from limited measurements [26]. Meanwhile and thereafter, there have been different deep learning techniques developed for learning reconstruction from limited measurements.

This survey aims to review the papers from 2016- June 2020 in accelerating MR imaging with deep learning techniques for undersampled MR image reconstructions. And we will also discuss merits, limitations and challenges associated with such methods. Last but not least, this paper will provide a starting

T This research was partly supported by the National Natural Science Foundation of China (61871371, 81830056), Key-Area Research and Development Program of GuangDong Province (2018B010109009), the Basic Research Program of Shenzhen (JCYJ20180507182400762), Youth Innovation Promotion Association Program of Chinese Academy of Sciences (2019351).

Shanshan Wang, Taohui Xiao and Hairong Zheng are with Shenzhen Institutes of Advanced Technology, Chinese Academy of Sciences Shenzhen, 1068 Xueyuan Avenue, Shenzhen University Town, Guangdong, China (e-mail: sophiasswang@hotmail.com; ss.wang@siat.ac.cn; th.xiao@siat.ac.cn; hr.zheng@siat.ac.cn). Qiegen Liu is with the Department of Electronic Information Engineering, Nanchang University, Nanchang, China. (e-mail: liuqiegen@hotmail.com).

point for researchers interested in contributing to this field by pointing out good tutorial resources, state-of-the-art open-source codes and meaningful data sources.

## II. DEEP LEARNING BASICS

Deep learning uses huge neural networks with many layers of processing units, taking advantage of advances in computing power and improved training techniques to learn complex patterns in large amounts of data. It is a branch of machine learning and artificial intelligence. The most popular deep learning examples include convolutional neural network (CNN), Multilayer perceptron (MLP), U-Net, ResNet, Generative adversarial network (GAN), cycle GAN, Recurrent neural network (RNN) and so on so forth. The deep learning techniques should use feedforward neural networks. There are different connections and components needed for designing the network.

### A. Convolutional layer

Instead of using general matrix multiplication, deep learning networks especially CNN uses convolution operations to extract features. Specifically, convolution is a mathematical operation that is the integral of the product of the two inputs with one reversed and shifted. In the network, the activations are convolved with a set of small parameterized filters, collected in a tensor W. Each filter shares the same weights across the whole input domain. This weight-sharing strategy can help reduce the size of the needed parameters and is motivated by the fact that similar structures and features exist in the image. Appling all the convolutional filters at all locations of the input to a convolutional layer produces a tensor of feature maps.

### B. Activation layer

The features extracted from the convolutional layer are fed through activation functions, known as activation layer in the network. Nonlinearity is an important characteristics of deep neural networks, which is generated using nonlinear activation function. Popular nonlinear functions include sigmoid or logistic, hyperbolic tangent (Tanh), rectified linear units (ReLU) and so on so forth. The activation function is to make the network more powerful and add ability to the network to learn complex functional mappings between inputs and outputs. Besides nolinearity, differentiability is another character of the activation function.

### C. Pooling

This is a key technique in current deep learning technologies. Pooling divides the input map into a set of rectangles and outs a single value for each rectangle. The operation is normally computed by average or max functions or sum functions. Since small variances in images result in small changes in the activation maps, the pooling layers give the neural network certain degree of translational invariance capability. Pooling has the downsampling effect. An alternative way to achieve this downsampling effect is the convolution with increased stride.

### D. Fully-connected layer

Fully connected layer is the most popular connection for traditional multilayer perceptron and have also been frequently used in the current deep learning techniques. It means that the neuron of each layer is connected to every neuron in the previous layer. Adding a fully-connected layer is also a cheap way of adding nonlinearity to the deep learning architectures.

### E. Batch normalization

Batch Normalization (BN) can adjust and scale the activations of the network layers. It has quite a few encouraging properties. First, it normalizes the distribution of input data and can accelerate the model learning speed. Second, it can reduce the sensitivity to weight initialization and scale, and therefore simplifies the tuning process and makes network learning more stable. Third, BN can reduce overfitting with its regularization effect.

### F. Dropout regularization

Randomly removing neurons during training is known as dropout regularization. It is often regarded as an averaging technique from the view of stochastic sampling of neural networks. With the dropout employed, slightly different networks are obtained for each batch of training data.

### G. Capacity, overfitting and under-fitting

Capacity, overfitting and under-fitting are basic concepts often encountered in artificial intelligence. Capacity generally refers to the capability of a network or a model in representing a relationship. It is normally proportional to its network size. The deeper and the wider, the model capacity will be larger. Overfitting means that the networks have overfit the training data where the gap between training error and test error is too large. Overfitting tends to extract some of the residual variations as the model structures. Under-fitting means that the model cannot obtain a sufficiently low error on the training set, which cannot capture the underlying structure of the data. When we configure our models, we need to consider about the correlation between model capacity, the problem complexity, and data size and heterogeneity. The data size is not the larger the better. The model capacity isn't the lager the better either. When the data size is complex and model capacity is small, underfittng may happen. On the other hand, if the model capacity is large and data size is small, overfitting will happen. So we need to properly adjust the model capacity and training data size for achieving an appropriate performance for the image reconstruction problem.

### H. Backward propagation

It is a method to update the weights of the network by calculating the gradient of each layer with respect to the loss function. Backpropagation is shorthand for the "backward propagation of errors" since the error is computed at the output and distributed backwards through the network layer.

## III. MR IMAGE RECONSTRUCTION BASICS

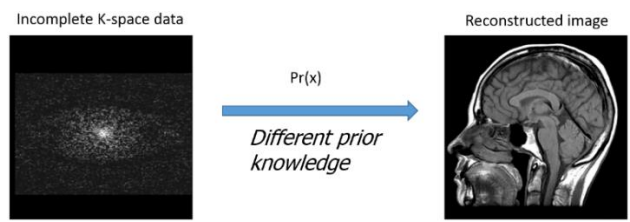

Fig. 1 Image reconstruction from incomplete k-space data

For MR imaging, the original data obtained by MR scanners can be modelled as follows

$$y = Ex + n \quad (1)$$

where y is the K-space measurements. x means the image to be reconstructed. E represents the encoding matrix for MR imaging; It could be E=MF for single channel imaging or MFS for the parallel imaging scenarios. M is the sampling mask; F represents Fourier transform and S is the sensitivity encoding; n is the noise, which was introduced as the disturbance in the measurement process.

If the sampling data y satisfies the Nyquist sampling theorem, the image can be directly updated. For example, if the data is fully sampled in a Cartesian grid, direct inverse Fourier transform can be used. If the data is subsampled, different regularization like sparsity, low-rankness constraints should be employed to attack the under-determined property of the subsampling. Compressed sensing (CS) is one of the revolutionary approaches for solving the problem. The image reconstruction problem normally can be described as follows with the promotion of the optimized signal-to-noise ratio (SNR),

$$f(x) = \arg\min_x \frac{1}{2}\|Ex - y\|_2^2 + \lambda P_r(x) \quad (2)$$

where the first term means data-fidelity and the second one $P_r(x)$ denotes prior regularization. λ means a weighting parameter, which determines whether or not prior information is introduced and how much the prior information contributes to the final reconstruction. The prior information could be fixed transforms such as wavelet, total variation and sigular value decomposition ones, or adaptive ones such as dictionary learning, data-driven tight frames and so on so forth [27]–[36].

## IV. DEEP LEARNING FOR MR RECONSTRUCTION

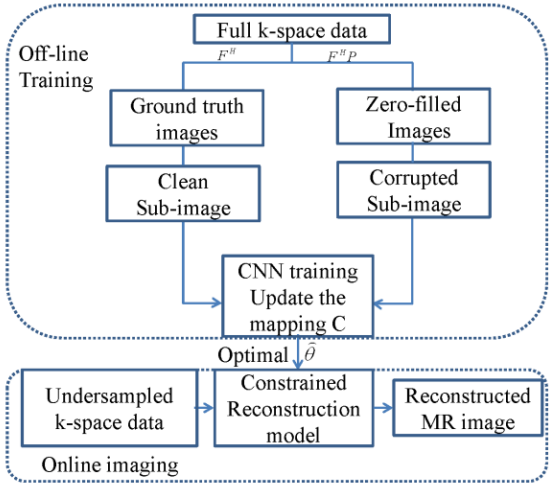

Fig.2 The deep learning MR reconstruction framework [26]

In 2016, [26] introduces deep learning into fast MR imaging which uses offline trained models to achieve accurate online MR reconstruction. During the offline training step, the model parameters are optimized through minimizing a loss function calculated between the reconstructed and the reference images. During the online reconstruction, the optimized model is adopted to generate a network prediction, which can be used for regularization or direct reconstruction

In the meantime and thereafter there are more and more methods. Based on the reconstruction framework and its inputs and outputs, these methods can be roughly categorized into two types, data-driven end-to-end deep learning MR image reconstruction and physics/model-driven unrolling iterative deep learning methods. The former tries to learn a nonlinear mapping between the data pairs of aliased images/incomplete k-space to the artifact free/full k-space. The physics/model-driven methods try to solve an inverse problem with the help of traditional iterative algorithms.

### A. Data driven end-to-end deep learning MR image reconstruction

As shown in Fig. 3, based on the input and output data pairs, the data-driven end-to-end learning reconstruction methods can be further categorized into three subtypes. 1) the image domain learning between the aliased images and high quality MR images; 2) the k-space to image space hybrid learning between the k-space and high quality MR images; 3) k-space domain learning which explores the k-space correlations. Data-driven end-to-end learning reconstruction benefits a lot from the deep learning models developed for natural image processing. Different types of learning reconstructions have different kinds of properties. Generally, the image domain end-to-end learning are better at removing image noise and artifacts, which can directly use different network structures and transfer learning techniques. The k-space learning are better at keeping high frequency information namely details and fine structures, which has strong connections with classical K-space reconstruction methods. The K-space to image space learning can get a better tradeoff between removing noise and artifacts, and keeping details. We present some main network architectures used for the data-driven end-to-end deep learning MR image reconstruction.

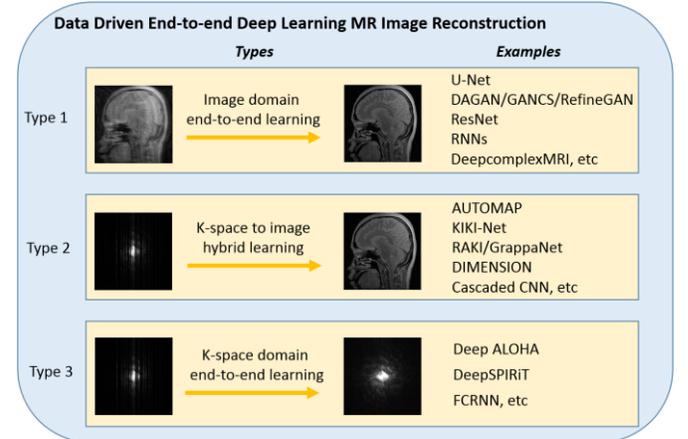

Fig.3 End-to-end deep learning MR image reconstruction: directly learn a nonlinear mapping between the data pairs of aliased images/incomplete k-space to the artifact free/full k-space. The figure presents three subtypes with their corresponding examples.

*1) U-Net*

U-Net was originally proposed to be used for biomedical image segmentation, and achieved excellent segmentation results. It contains the encoder structure on the left and the decoder structure on the right. A total of four downsampling and four upsamplings form a U-shaped structure. U-Net uses a skip-connection connection. This step merges the location information of low-level features with the semantic information of deep-level features, which promotes the network to get better results. The difference between U-Net and autoencoder is that U-Net has a skip-connection structure. At present, many studies have used U-Net structure or as a sub-structure, and have achieved effective results. In the study [37]–[40], the author designed a multi-resolution deep convolutional framelets based on U-Net to recover high-resolution MR images from undersampled K-space data, and combined residual learning to further improve the reconstruction performance of the network. The author in Study [41] proposed a fast and accurate deep learning reconstruction method for human lung gas MRI, which consists of coarse-to-fine nets (C-net and F-net). The loss functions used in C-net and F-net training are L2 loss and united L2 loss with proton prior knowledge, and the core networks of C-net and F-net use U-Net. Compared with traditional CS-MRI, the proposed deep learning method can better reconstruct the human lung gas MR images acquired from high-undersampling k-space. The author in Study [42] proposed a deep learning undersampling MRI reconstruction method using U-Net and k-space correction. This method shows remarkable performance, and only 29% of k-space data can effectively generate high-quality MR images. In research [43], iterative learning based on U-Net is used for compressed sensing MRI reconstruction. Compared with the original end-to-end U-Net, the iterative learning strategy adopted has further improved the reconstruction performance. In addition, there is a series of works using U-Net or improved networks based on U-Net for fast MRI reconstruction studies [37], [44]–[48].

*2) Generative adversarial network (GAN)*

Ian Goodfellow et al. first proposed GAN for the generation of natural images [49]. GAN consists of two networks, one is a generator and the other is a discriminator. Inspired by the zero-sum game, these two networks confront each other to achieve the best generation effect. The role of the generator is to generate images that are close to the real image after training. The role of the discriminator is to accurately identify whether the image is generated or real. The two networks are continuously trained, and finally the discriminator cannot distinguish the generated Image and real image. After continuous development of GAN, different types of versions have been derived [50], such as Vanilla GAN, CGAN, DCGAN, GRAN, LAPGAN, WGAN, etc. Research [51] used GAN for the reconstruction of MRI images.

Research [52] also applied GAN in the field of rapid MRI, using the adversarial neural network to estimate unsampled data. For 1.5T MRI using only 52% of the original data, promising image reconstruction results were obtained. DAGAN was proposed in the study [53] for fast compressed sensing MRI reconstruction. In order to stabilize the U-Net-based generator, they designed a refinement learning method. The author combines adversarial loss with innovative content loss to better preserve texture and edge information during image reconstruction, taking into account frequency information. Compared with the traditional CS-MRI reconstruction method and the latest deep learning method, the reconstruction result of DAGAN is better. The author in study [54] proposed a new deep learning based confrontation model RefineGAN for fast and accurate CS-MRI reconstruction. RefineGAN is an improved model based on fully residual convolutional autoencoder and GANs, employing deeper generator and discriminator with cyclic data consistency loss, so as to achieve the purpose of faithful interpolation for a given under-sampled k-space. RefineGAN is superior to the state-of-the-art CS-MRI method in terms of efficiency and image quality. In addition to the above methods, there is a series of work that uses GAN for fast magnetic resonance image reconstruction [55]–[63].

*3) ResNet*

Residual network (ResNet) has made outstanding achievements in natural image classification since it was proposed. ResNet or its residual structure has been widely used in detection, segmentation, recognition and other application directions. Similarly, in the field of medical image reconstruction, ResNet or residual structure are also widely used. The author in study [64] proposed a deep learning MR image super-resolution reconstruction network based on residual learning. The network uses both global residual learning (GRL) and local residual learning (LRL) to improve image reconstruction performance. Research [65] directly used ResNet to reconstruct T2 mapping from overlapping-echo detachment (OLED) sequence. An effective reconstruction effect has been achieved on simulated MR images with single-shot OLED sequence. Research [66] proposed a deep ResNet using variable density spiral trajectory for faster and better fMRI reconstruction. The deep residual network consists of various residual blocks. An enhanced recursive residual network (ERRN) based on recursive residual network was proposed in the study [67] for under-sampling MR image reconstruction and improving the image reconstruction quality. The authors in study [68] proposed a residual CNN for super-resolution reconstruction of single anisotropic 3D MR images based on residual learning. The proposed residual CNN with long and short skip connections can effectively restore the high-frequency details of MR images. In addition, residual networks or residual learning are also used in studies [69], [70] and the aforementioned studies [37]–[40].

*4) Complex-valued neural network*

Initially, deep learning MRI mostly used only the amplitude information of MR images. However, considering the complex nature of MR images, the phase also contains important information that cannot be ignored in MR reconstruction [71]. Many subsequent studies not only use amplitude information but also phase information, or divide MR complex data into real and imaginary parts for training. Complex-valued neural nets with 1-channel complex MRI signals for MRI fingerprinting have proved to be better than real-valued networks of complex MRI signals represented with 2-channel real/imaginary input [72]. The study in [73] proposed the Complex Dense Fully Convolutional Network (CDFNet), which uses densely connected fully convolutional blocks to support deep learning operations on complex-valued data. The author in [74] proposed Deepcomplex MRI, a deep residual convolutional neural network that considers the correlation between the real and imaginary parts of MR complex images, and includes the k-space data consistency. Deepcomplex MRI has achieved better results than real-value networks, traditional CS-MRI, and the most advanced deep learning reconstruction algorithms. In addition, the research [37] mentioned in the previous U-Net chapter and the KIKI-net [75] and DIMENSION [76] in the subsequent Dual-Domain Neural Network chapter divide the input data into real and imaginary parts for training. The author in [39] divided the complex-valued data into magnitude and phase and used two networks for training. The research in [71], [77] also directly uses complex convolutional neural networks to reconstruct MR images.

*5) Dual-domain neural network*

MR scanned data is collected in the Fourier domain (called k-space), and the acquisition time is proportional to the amount of k-space data collected. So fast MR imaging research can be based on k-space or image domain. At present, the end-to-end deep learning magnetic resonance imaging research mostly starts from the image domain, such as the aforementioned GAN, U-Net, etc, and a small part starts from k-space, such as Automated Transform by Manifold Approximation (AUTOMAP) [78]. However, there are still some studies that use the frequency domain and the image domain at the same time, which we call the dual-domain neural network. The author in [75] proposed KIKI-net, which is a cross-domain CNN. KIKI-net uses a deep CNN to process k-space data (KCNN) and another network to process image domain data (ICNN), and embeds data consistency operations. These components are alternately applied. KIKI-net can effectively restore the MR image tissue structures and remove aliasing artifacts. The author in [76] proposed a multi-supervised network training method for dynamic MR imaging that integrates a priori knowledge in the k-space and spatial domain, called DIMENSION. DIMENSION consists of two parts, one is the frequency domain network used to update the K-space information (FDN), and the other is the spatial domain network used to extract the high-level features (SDN) of MR images. In dynamic MR imaging, DIMENSION achieves better reconstruction results in a shorter time, compared with current state-of-the-art methods. MD-Recon-Net and DD-DLN were proposed in [79], [80] for compressed sensing MRI. Both of these networks take into account both the frequency domain and image domain information, and achieve a good visual effect. In addition, there are other similar deep network MRI reconstruction studies that use prior knowledge in both the frequency domain and the image domain [81]–[83].

*6) Recurrent neural networks (RNNs)*

RNNs are a type of neural network that uses sequence information to process a series of inputs. The difference between RNNs and other neural networks is that the nodes between hidden layers are connected. The input at each moment not only has the input at the current moment, but also the output value of the hidden layer at the previous moment. This allows RNNs to learn more historical information and use it in current predictions. This kind of structure makes this type of network have a memory function, so it is widely used in data related to time series. In the field of rapid magnetic resonance imaging research, there are studies using RNNs for magnetic resonance image reconstruction. The author in [84] proposes a novel end-to-end method for MR image reconstruction based on RNN. The RNN in this paper extracts features for image reconstruction and simultaneously sweeps the k-space both horizontally and vertically. The author in [85] uses self-supervised RNN for 4D Abdominal and In-utero MR Imaging.

It firstly uses self-supervised RNN for respiratory motion estimation, and then use a 3D deconvolution network for super-resolution reconstruction. Research [86] jointly explores the dependencies of temporal sequences and the iterative characteristics of traditional algorithms. It uses the bidirectional RNN across time sequences to learn the spatio-temporal dependencies. In addition, the author in [87] proposed sRAKI-RNN based on RAKI to accelerate MR imaging.

*7) Hybrid architecture models*

There are also hybrid network architecture models, which usually integrates different structures. Reference [88] proposes a multi-modal fusion deep learning fast magnetic resonance reconstruction method. The deep learning method proposed in this paper combines dense block and U-Net. Reference [89] proposed a SANTIS method for accelerating magnetic resonance image reconstruction. SANTIS combines data cycle–consistent adversarial network, end-to-end convolutional neural network mapping, data fidelity enforcement, and sampling-augmented training strategy. The deep learning network used combines U-Net and residual structure and GAN. Reference [90] proposes a self-attention convolutional neural network, which contains self-attention module, U-Net, residual module and dense connection. The self-attention CNN proposed in the article can improve the quality of the reconstructed image. In addition, in the studies [37]–[40], [45], [61], [91]–[93], a hybrid architecture containing several different network structures (such as GAN, Automap, U-Net, attention, dense block, residual learning, etc.) were used for fast magnetic resonance imaging research. The hybrid architecture models used in these references have achieved encouraging performances.

*8) Other CNN variants reconstruction models*

In addition to the above methods, we have further searched and sorted out additional related works using pure network methods. AUTOMAP provides a data-driven supervised learning reconstruction method with manifold learning [78], where the input of the network is sensor-domain-sampled data, and the output is the reconstructed image. And the fully connected layer is interspersed in the model structure, which can effectively reconstruct the MR image. In addition, AUTOMAP can not only reconstruct sensor data sampled by Cartesian, but also directly reconstruct the non-Cartesian samples, which promote the development of new data acquisition strategies across imaging modalities [78]. There are studies using CNN to perform real-time MR reconstruction on a single patient [94]–[96]. Deep attention networks have been used to study super-resolution magnetic resonance imaging reconstruction [97]. There is a robust artificial neural-networks for k-space interpolation (RAKI) method that directly uses deep learning algorithms for k-space interpolation to complete k-space data reconstruction [98]. This is a deep learning extension of the traditional GRAPPA algorithm. GRAPPA uses the fully sampled center of k-space and interpolation to estimate unsampled k-space lines [20]. RAKI uses CNNs trained from the fully sampled k-space center as interpolation functions to fill the unsampled k-space. Besides, deep convolutional neural networks based reference-driven compressed sensing MR image reconstruction was studied in [99]. Deep multistream CNN for parallel imaging in TOF magnetic resonance angiography was studied in [100]. [24] investigates the influence of network structure and loss functions on MR image reconstruction based on U-Net and Resnet. There are also studies on the impact of undersampling pattern optimization [101]. The authors in [102], [103] used deconvolution deep neural networks or multi-scale information fusion CNN for medical image super-resolution reconstruction. And the authors in [104] used Bayesian deep learning to accelerate magnetic resonance image reconstruction. In addition, there are some other studies on the use of convolutional neural networks for magnetic resonance image reconstruction [27], [105]–[130].

### B. *Physics/Model based unrolling iterative deep learning methods*

Different from the data-driven approaches, which typically need large datasets and re-training when acquisition parameters such as undersampling patterns are modified. The physics/model driven methods take utilizations of the MRI physics knowledges to solve an inversed problem with regularized prior knowledge. Here are some typical examples of physics/model driven learning MR reconstruction methods.

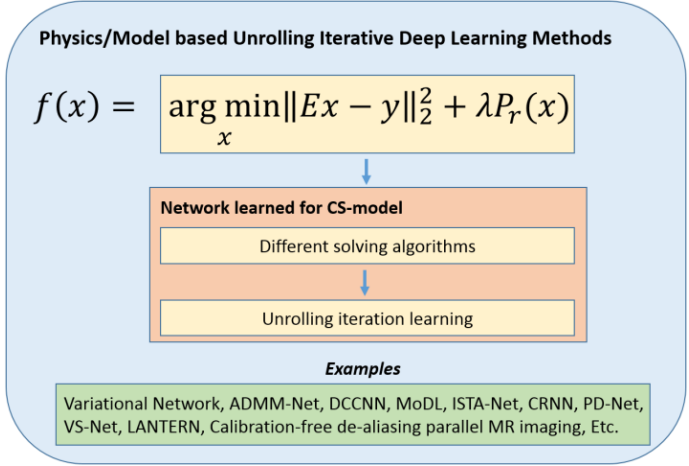

Fig.4 The figure presents the general process of model-based unrolling iterative deep learning MR image reconstruction with corresponding examples (In the green box).

*1) Variational Network (VN)*

The variational network (VN) [131] is proposed for accelerated and high-quality reconstruction of multi-channel MRI images. It is composed of a variational model based on generalized compressed sensing reconstruction formulation and deep learning, and uses an unrolled gradient descent scheme to learn all the parameters in the formulation. In variational network, the solution to (2) is

$$x^{n+1} = x^n - \sum_{i=1}^{N_k}(K_i^n)^T \Phi_i^{n'}(K_i^n x^n) - \lambda^n A^*(Ax^n - f) \quad (3)$$

where $K_i^n$ can be modeled as a convolution, $\Phi_i^{n'}$ is the non-linear activation function, $\lambda^n$ is the data term weights, $A^*$ is the adjoint operator, the sensitivity maps are used in the operators $A, A^*$, and $0 \leq n \leq T - 1$.

The network structure of VN is obtained by unfolding the iterations of Equation (3). The input of the network includes: the undersampled k-space data, coil sensitivity maps and the zero filling solution. In the training process of VN, the filter kernels $K_i^n$, activation functions $\Phi_i^{n'}$ and data term weights $\lambda^n$ are learned. And the complex-value images are divided into the real and the imaginary filters to learn separately, and $\Phi_i^{n'}$ combines the filter responses of these two feature planes.

The learning-based VN reconstruction method is superior to traditional reconstruction methods in a wide range of pathologies and provides a faster reconstruction speed, which is of great significance for integration into clinical workflow. In addition, studies [132], [133] carried out further research on accelerated MR image reconstruction based on VN.

*2) ADMM-Net*

The alternating direction method of multipliers network (ADMM-Net) was first proposed in research [134] for compressed sensing MRI, which uses the unrolling ADMM algorithm to learn the regularization parameters in CS-MRI. In ADMM-Net, the optimization problem of Formulation (1) can be solved efficiently based on ADMM algorithm. By introducing auxiliary variables $z = \{z_1, z_1, ..., z_L\}$ and a transform matrix $D_l$ for a filtering operation, the ADMM iterations can be written as the following three subproblems:

$$\begin{cases} x^{n+1} = \arg\min_x \frac{1}{2}\|Ax - y\|_2^2 - \sum_{l=1}^L \langle \alpha_l^n, z_l^n - D_l x \rangle \\ \qquad\qquad + \sum_{l=1}^L \frac{\rho_l}{2}\|z_l^n - D_l x\|_2^2, \\ z^{n+1} = \arg\min_z \sum_{l=1}^L \lambda_l g(z_l) - \sum_{l=1}^L \langle \alpha_l^n, z_l^n - D_l x^{n+1} \rangle \\ \qquad\qquad + \sum_{l=1}^L \frac{\rho_l}{2}\|z_l^n - D_l x^{n+1}\|_2^2, \\ \alpha^{n+1} = \arg\min_\alpha \sum_{l=1}^L \langle \alpha_l^n, D_l x^{n+1} - z_l^{n+1} \rangle, \end{cases} \quad (4)$$

where $n \in [1,2,...,N_s]$ means n-th iteration, $g(\cdot)$ is a general regularization function, and $z_l = D_l x$. The purpose of the transform matrix $D_l$ is to make the $z_L$ sparse through sparse images.

Then the solution of these subproblems is:

$$\begin{cases} X^n: x^n = F^T[P^T P + \sum_{l=1}^L \rho_l F D_l^T D_l F^T]^{-1} \\ \qquad\qquad [P^T y + \sum_{l=1}^L \rho_l F D_l^T (z_l^{n-1} - \beta_l^{n-1})], \\ Z^n: z_l^n = S(D_l x^n + \beta_l^{n-1}; \lambda_l/\rho_l), \\ M^n: \beta_l^n = \beta_l^{n-1} + \eta_l(D_l x^n - z_l^n), \end{cases} \quad (5)$$

where $x^n$ can be computed by fast Fourier transform, $S(\cdot)$ is a nonlinear shrinkage function for $g(\cdot)$, and the parameter $\eta_l$ is an update rate.

The author is the first to map the ADMM algorithm to a learnable deep learning architecture, achieving high reconstruction accuracy in MR images. The author then conducted an extended study based on ADMM-Net, and proposed Generic-ADMM-CSNet [108] for CS imaging tasks by re-designing and unrolling the ADMM algorithm, which further improved the reconstruction performance of this type of method. In addition, there are some other work using ADMM-Net for CS-MRI reconstruction [135].

*3) Cascaded convolutional neural network*

In the study [136], [137], the author proposed a deep network structure that cascades convolutional neural networks to accelerate the acquisition speed of MR image data, and the study uses the cardiac MR dataset. The cascaded CNN consists of CNN and the data consistency (DC) layer and the data sharing (DS) layer, where DS layer is used in the reconstruction of dynamic sequences. The principle of the cascaded CNN proposed by the author is that the output of the previous CNN is connected to a new CNN, thereby establishing a deep network that iterates between intermediate de-aliasing and DC layer reconstruction [136]. The authors show that when reconstructing each 2D image frame independently, the reconstruction performance of the cascaded CNN is significantly better than the state-of-the-art 2-D compressed sensing method. When reconstructing dynamic image sequence frames, the authors prove that combining 3D convolution and DS layer can effectively learn the spatio-temporal correlation of the input image and promote network reconstruction performance. Cascaded Dilated Dense Network (CDDN) was proposed for MRI reconstruction. The CDDN is composed of multiple sub-network iterations [138]. Each sub-network contains a normal convolution layer, dense blocks with residual connection, dilated convolution, and De-Aliase Module. After each sub-network, a two-step Data Consistency (TDC) operation is performed on the k-space. CDDN has achieved effective test results on both the cardiac dataset and the FastMRI [139] dataset. In addition, the research [82], [140] also adopted the cascade network structure.

*4) MoDL*

Model-based reconstruction using Deep Learning prior (MoDL) architecture is proposed to solve general inverse problems. The author [141] uses a variational framework containing data consistency items and learned CNN to capture the redundant information of the image, and unrolls it into a deep network based on the alternating recursive algorithm. The author uses the following MRI image reconstruction constraint formula that is the same as formula (2).

$$x_{rec} = \arg\min_x \|\mathcal{A}(x) - b\|_2^2 + \lambda \|\mathcal{N}_w(x)\|^2 \quad (6)$$

where $\|\mathcal{A}(x) - b\|_2^2$ is a data consistency item, $\|\mathcal{N}_w(x)\|^2$ is a regularization prior that uses a learned CNN to estimate noise and alias patterns, $\lambda$ is a trainable regularization parameter.

After Taylor series and a series of approximations, the solution of equation (6) is

$$x_{n+1} = (\mathcal{A}^H\mathcal{A} + \lambda I)^{-1}(\mathcal{A}^H(b) + \lambda z_n) \quad (7)$$
$$z_n = D_w(x_n) \quad (8)$$

where $D_w(x)$ is the "denoised" version of $x$ that removes noise and alias artifacts.

Compared to direct inversion approaches, MoDL explicitly considers the forward model, using a smaller network with fewer parameters to capture image information well. Therefore, the demand for network training data and training time is reduced. In addition, $D_w$ weight sharing strategy in each iteration improves the performance of the model compared to methods that rely on pretrained denoisers. After that, the author carried out further research on dynamic MRI based on MoDL [142].

*5) ISTA-Net*

Inspired by the Iterative Shrinkage-Thresholding Algorithm (ISTA) used to optimize the regularization term of the CS reconstruction model, ISTA network (ISTA-Net) [143] was developed. The author proposes an effective strategy to use nonlinear transformation to solve the proximal mapping related to sparsity-inducing regularizer, thereby effectively transforming ISTA into a deep network form.

In ISTA-Net, as an unrolling iterative version of traditional ISTA, through the use of a general form of image transformation F(x), the optimization problem of Equation (2) is written as:

$$r^{n+1} = x^n - \rho \Phi^T(\Phi x^n - y) \quad (9)$$
$$x^{n+1} = \arg\min_x \frac{1}{2}\|F(x) - F(r^{n+1})\|_2^2 + \theta\|F(x)\|_1 \quad (10)$$

where $\rho$ is the step size, $k$ is the ISTA iteration index, $\theta$ is the merged parameter related to regularization parameter $\lambda$ of Equation (1) and the parameters of F(·). The output image can be updated as:

$$x^n = \tilde{F}^n(soft(F^n(\gamma^n), \theta^n)) \quad (11)$$

where $soft(\cdot)$ is soft thresholding, $\tilde{F}(\cdot)$ is the left inverse of F(·).

During ISTA-Net training, there are a total of $N_p$ phase iterations, and each phase iteration corresponds to one iteration in ISTA. The author also proposed an enhanced version of ISTA-Net, called ISTA-Net$^+$, which can further improve the performance of CS reconstruction. In addition, studies [144] also unrolled the ISTA algorithm into a deep network structure and applied it to MRI reconstruction.

*6) PD-Net*

Research [145] uses primal-dual hybrid gradient (PDHG) algorithm and adopts an iterative scheme to solve ill-posed inverse problems. The form of the optimization problem is:

$$\min_{f \in X}[F(\mathcal{K}(f)) + \mathcal{G}(f)] \quad (12)$$

where $\mathcal{K}: X \to U$ is a (possibly non-linear) operator, $U$ is a Hilbert space, $F: X \to \mathbb{R}$ and $\mathcal{G}: X \to \mathbb{R}$ are functionals on the dual/primal spaces. The solution of Equation (11) can be obtained by the PDHG algorithm.

Further, a learnable primal-dual algorithm [146] is proposed for tomographic reconstruction, which is an algorithm combining deep neural network and model-based reconstruction by unrolling iterative a proximal primal-dual optimization problem. Currently, the PDHG algorithm has been widely applied to CT [123].

Similarly, if $F(\mathcal{K}(f)) = \frac{1}{2}\|Ax - y\|_2^2$, we can regard equation (2) as a special form of equation (12). Therefore, an unrolled iterative version of the PDHG algorithm was proposed for accelerated MR imaging, named PD-Net [147]. In PD-Net, parameterized operators are used instead of the proximal operators, and the parameters inside are automatically learned through offline training. Use the following formula to unrolled iterations for primal updating $x_{n+1}$ and dual updating $z_{n+1}$:

$$\begin{cases} x_{n+1} = \mathcal{L}(x_n, Ax_n, f) \\ z_{n+1} = \mathcal{T}(z_n, A^*x_{n+1}) \end{cases} \quad (13)$$

Compared with the traditional CS reconstruction method and other unfolding iterative methods, PD-Net has achieved better reconstruction results.

*7) Other unrolling optimization models*

In addition to the several typical unrolling iterative methods introduced above, there are also some other unrolling iterative MRI reconstruction methods. Convolutional recurrent neural networks (CRNN) [86] are inspired by variable splitting and alternate minimization strategies, and are proposed for cardiac MR image reconstruction by jointly exploring the dependencies of spatial sequences and the iterative characteristics of traditional optimization algorithms. The author in the study [148] proposed a model based convolutional de-aliasing network to accelerate parallel MR imaging with adaptive parameter learning exploring both spatial redundancy and multi-coil correlations. Unlike most existing parallel reconstruction algorithms, the de-aliasing reconstruction model can perform fast MRI reconstruction from highly undersampled k-space data without explicit sensitivity calculation.

Subsequently, the author further proposed the learn analysis transform network for dynamic MR imaging (LANTERN) [149]. LANTERN integrates CS-MRI iterative optimization model and deep learning, using adaptively trained CNN sparse constrained spatial and temporal domains information, and using small data sets can achieve excellent reconstruction performance. Variable splitting network (VS-Net) [150] was proposed to accelerate parallel MR image reconstruction by unrolling the resulting iterative process of a variable splitting optimization. Each iterative reconstruction process of VS-Net contains three modules: denoiser block (DB), data consistency block (DCB), and weighted average block (WAB), which achieve better results than the state-of-the-art method under the same conditions. In addition, there are some other fast MRI research work that also uses unrolling iterative reconstruction methods, such as related work in [151]–[154].

## V. OPEN SCIENCE AND REPRODUCIBLE WORKS

Deep learning for MR imaging is a very dynamic research area. Luckily, there are some researchers posting their codes on the GitHub platform https://github.com or their own website. The datasets used are normally available through various repositories. If a researcher is interested in this topic, one could find an implementation uploaded to Github and a method described in a paper. This forms a good starting point for the beginners. We have surveyed the datasets, code website and summarized them in the table 1.

## VI. CHALLENGES, LIMITATIONS AND FUTURE PERSPECTIVES

It is clear that deep learning has introduced new opportunities. As a first discussion, we summarize the pros and cons of different learning reconstruction. Generally, the image domain end-to-end learning are better at removing image noise and artifacts, which can directly use different network structures and transfer learning techniques. But the image may tend to a little blurry due to strong denoising effect of the network. The k-space learning are better at keeping high frequency information namely details and fine structures, which has strong connections with classical K-space reconstruction methods. However, noise or artifacts sometimes tend to appear. The K-space to image space learning can get a better tradeoff between removing noise and artifacts, and keeping details. Comparing the data-driven approaches with the physics/model based approaches. The data-driven ones normally need more training data and the networks used are more complicated. In terms of speed, end-to-end learning requires a long training time, and the training time for the unrolling iterative method is relatively short. In terms of robustness, when there is an adversarial sample or an adversarial attack, the unrolling iterative method shows relatively more robust reconstruction performance. In terms of interpretability, physics based unrolling method are more explainable. The end-to-end method learns the mapping relationship between data, whose intermediate process is a relatively black box and has poor interpretability. The unrolling iteration method is based on the traditional CS model and has a series of mathematical derivation processes, which is more interpretable.

Secondly, we discuss if the techniques depend on the fully sampled dataset. At present, most end-to-end deep learning MRI methods are supervised learning, such as AUTOMAP, KIKI-Net, DIMENSION, DeepcomplexMRI, etc. These networks require training labels. Semi-supervised learning generally refers to the form of training that contains only part of the label data. These works are mainly based on GAN. Weak-supervised learning uses weak labels [155] generated by other methods. There are also some unsupervised learning methods, such as CycleGAN or self-supervised learning. At present, some studies have adopted self-supervision [85], [156], [157]. The self-supervised learning method still has great potentials since it does not require training labels and can complete the corresponding task only from the data itself. It is one of the important future research directions.

Finally, we rethink about the MRI workflow. The current MRI workflow is generally from data acquisition to image reconstruction and then to image analysis and diagnosis. The workflow may change with the development of artificial intelligence techniques. For example, the SegNetMRI proposed in the study [158] can achieve simultaneous image reconstruction and segmentation. The Joint-FR-Net proposed in the research [159] obtains the image segmentation results directly from the k-space data. These methods are all directly from the original k-space data. We believe that more research works will appear in the future, not only from k-space to segmentation, but also from k-space to classification, detection and even diagnosis. The task-driven MR reconstruction works may become more and more popular.

Opportunity always lies in Challenges. We may need to devote endeavors to answer many questions, e.g. how many data are necessary; what's the optimal network architecture; how many layers are in need; what's the configurations for different applications and theoretical supports are in need as well. As an outlook, we think mathematicians, physicians and imaging scientists will work more closely. New trends in machine learning in MR reconstruction, may include training with small data sets, more weakly supervised or unsupervised learning works, network development to address high-dimensional imaging and multitasking, etc.

## VII. CONCLUSIONS

This article surveyed the papers in deep learning for MR image reconstructions from incomplete k-space data, which has shown big potential for the next generation of fast MR imaging technique with promising performances achieved. The data-driven method directly learns the mapping relationship between under-sampled data and fully-sampled data, so as to achieve high quality reconstruction. The unrolling iterative method combines the traditional physics models with deep learning. In general, these approaches can get better reconstruction results and higher acceleration factors with strong prior knowledge learning capabilities. While most of the methods show encouraging performances and point out interesting directions, there are also some misconceptions happening from time to time and ignorance of the data bias and domain shift issues, which care must be given to, so as to fight for a brighter future. Limitations include the requirement of large amount of training data; the prior knowledge exploited may be constrained to the data/artifacts seen during training; the theoretical explanations are still underdeveloped. Opportunity always lies in Challenges.

As an outlook, we believe mathematicians, physicians and imaging scientists will work more closely in the future. New trends in machine learning in MR reconstruction, may include training with small data sets, weakly supervised/unsupervised learning, network development to address high-dimensional imaging and multitasking. It can be expected that more robust methods with strong theoretical explanations will be available if we continue to devote efforts in this direction.

Table 1. The surveyed deep learning fast magnetic resonance imaging research work with open source code or open source data

| Category | Reference | Year | Method | Network | Data | Code |
|---|---|---|---|---|---|---|
| Data Driven End-to-end Deep Learning MR Image Reconstruction | Zhu et al. [78] | 2018 | AUTOMAP | CNN | ImageNet database (http://www.image-net.org/), MGH-USC HCP public database (https://db.humanconnectome.org/) | https://github.com/chongduan/MRI-AUTOMAP |
| | Han et al. [37] | 2020 | — | U-Net | Knee k-space dataset (http://mridata.org/), MGH-USC HCP public database (https://db.humanconnectome.org/) | https://github.com/hanyoseob/k-space-deep-learning |
| | Quan et al. [54] | 2018 | RefineGAN | GAN | the IXI database [184] (the brain dataset) and the Data Science Bowl challenge [185] (the chest dataset) | http://hvcl.unist.ac.kr/RefineGAN/ |
| | Yang et al. [53] | 2018 | DAGAN | GAN | MICCAI 2013 grand challenge dataset, pathological MRI images | https://github.com/nebulaV/DAGAN |
| | Mardani et al. [51] | 2018 | GANCS | GAN | Abdominal Dataset; Knee Dataset | https://github.com/gongenhao/GANCS |
| | Cole et al. [77] | 2020 | Complex-valued CNN | U-Net | Knee images; body scans data; cine images | https://github.com/MRSRL/complex-networks-release |
| | EI-Rewaidy et al. [71] | 2020 | ℂNet | U-Net | Cardiac MR dataset | https://github.com/hossam-elrewaidy/urus-mri-recon |
| | Wang et al. [74] | 2020 | DeepcomplexMRI | ResNet | Brain dataset; Knee dataset (https://github.com/VLOGroup/mri-variationalnetwork) | https://github.com/CedricChing/DeepMRI |
| | Souza et al. [82] | 2019 | Hybrid Cascade model. | CNN | Calgary-Campinas brain MR raw data ((https://sites.google.com/view/calgary-campinas-dataset/home) | https://github.com/rmsouza01/CD-Deep-Cascade-MR-Reconstruction |
| | Zheng et al. [138] | 2019 | CDDNwithTDC | CNN | Cardiac real-valued MR images | https://github.com/tinyRattar/CSMRI_0325 |
| | Souza et al. [83] | 2019 | Hybrid-CS-Model-MRI | U-Net | Brain MR dataset (https://sites.google.com/view/calgary-campinas-dataset) | https://github.com/rmsouza01/Hybrid-CS-Model-MRI |
| | Wang et al. [76] | 2019 | DIMENSION | CNN | Cardiac MR data | https://github.com/Keziwen/DIMENSION |
| | Ran et al. [79] | 2020 | MD-Recon-Net | CNN | the Calgary-Campinas dataset (ttps://sites.google.com/view/calgary-campinas-dataset/home/mr-reconstruction-challenge) | https://github.com/Deep-Imaging-Group/MD-Recon-Net |
| | Huang et al. [92] | 2019 | MICCAN | CNN | Cardiac MRI dataset | https://github.com/charwing10/isbi2019miccan |
| | Akçakaya et al. [98] | 2018 | RAKI | CNN | Phantom imaging; In vivo imaging | https://people.ece.umn.edu/~akcakaya/RAKI.html |
| Physics/Model based Unrolling Iterative Deep Learning Methods | Yang et al. [134] | 2016 | ADMM-Net | CNN | Brain MR images from clinic and chest MR images (https://masi.vuse.vanderbilt.edu/workshop2013/index.php) | https://github.com/yangyan92/Deep-ADMM-Net |
| | Hammernik et al. [131] | 2018 | Variational network | CNN | Clinical knee dataset (https://github.com/VLOGroup/mri-variationalnetwork) | https://github.com/VLOGroup/mri-variationalnetwork |
| | Schlemper et al. [136] | 2018 | DCCNN | CNN | Cardiac MR dataset | https://github.com/js3611/Deep-MRI-Reconstruction |
| | Zhang et al. [143] | 2018 | ISTA-Net | CNN | Brain MR images | http://jianzhang.tech/projects/ISTA-Net. |
| | Aggarwal et al. [141] | 2019 | MoDL | CNN | Brain MR dataset | https://github.com/hkaggarwal/modl |
| | Yang et al. [108] | 2020 | ADMM-CSNet | CNN | Brain MR images from clinic and chest MR images (https://masi.vuse.vanderbilt.edu/workshop2013/index.php) | https://github.com/yangyan92/ADMM-CSNet |
| | Chen et al [148] | 2019 | Convolutional de-aliasing network | CNN | MR brain dataset | https://github.com/yanxiachen/ConvDe-AliasingNet. |
| | Duan et al. [150] | 2019 | VS-Net | CNN | Clinical knee dataset (https://github.com/VLOGroup/mri-variationalnetwork) | https://github.com/j-duan/VS-Net |
| | Qin et al. [86] | 2019 | CRNN | RNN | Cardiac cine MR images | https://github.com/js3611/Deep-MRI-Reconstruction |

ACKNOWLEDGEMENT

This research was partly supported by the National Natural Science Foundation of China (61871371, 81830056), Key-Area Research and Development Program of GuangDong Province (2018B010109009), the Basic Research Program of Shenzhen (JCYJ20180507182400762), Youth Innovation Promotion Association Program of Chinese Academy of Sciences (2019351).